\documentstyle[12pt]{article}
\begin{document}

\newtheorem{theorem}{Proposition}
\newtheorem{ex}{Example}

\begin{center}
{\Large{\bf Helicity invariants in 3D : \\[3mm]
kinematical aspects}} \\

\vspace{13mm}
{\large Hasan G\"{u}mral   \\
\vspace{5mm}
Feza G\"{u}rsey Institute  \\
P.O. Box 6, 81220 \c{C}engelk\"oy-\.Istanbul, Turkey} \\
\vspace{2mm}
hasan@gursey.gov.tr \\
\vspace{7mm}
\today
\end{center}

\vspace{1cm}

\section*{Abstract}

Exact, degenerate two-forms $\Theta_k=d\theta_k$ on time-extended
space $R \times M$ which are invariant under the unsteady,
incompressible fluid motion on three-dimensional region $M$ are introduced.
The equivalence class up to exact one-forms of $\theta_k$ is splitted by the
velocity field. The components of this splitting corresponds
to Lagrangian and Eulerian conservation laws for helicity densities.
These are expressed as the closure of three-forms $\theta_k \wedge \Theta_l$
which depend on two discrete and a continuous parameter.
Each $\Theta_k$ is extended to a symplectic form on $R \times M$.
The subclasses of $\theta_k$'s giving rise to Eulerian
helicity conservations is shown to result in conformally
symplectic structures on $R \times M$.
The connection between Lagrangian and Eulerian conservation
laws for helicity is shown to be the same as the conformal
equivalence of a Poisson bracket algebra to infinitely many
local Lie algebra of functions on $R \times M$.

\newpage

\section{Introduction}

In this work, we shall concern with the problem of constructing
infinitely many helicity type integrals for three dimensional
incompressible fluids analogous to the enstrophy
type Casimirs of two dimensional flows.
We shall express Lagrangian and Eulerian conservation laws
using invariant differential forms constructed for each
kinematical (particle relabelling) symmetry of the velocity field.
The relation between these two types of conservation laws
will be shown to be equivalent to the conformal relation
between Poisson and Jacobi structures on time-extended
space of flows.
The invariants under consideration are related to the description
of reduced phase space of Eulerian equations of ideal fluids
which are the orbits of coadjoint action of the group $Diff_{vol}(M)$ of
volume preserving diffeomorphisms of the flow domain $M$.

\subsection{The problem of coadjoint orbits}

The motion of an ideal fluid on a Riemannian manifold $M$ can be
formulated as geodesic motion with respect to a right-invariant
metric on $Diff_{vol}(M)$ \cite{arn66},\cite{mwe74}.
The Lie algebra  ${\cal X}_{div}(M)$
consists of divergence-free vector fields on $M$ tangential to
the boundary of $M$. The dual ${\cal X}^*_{div}(M)$ of the Lie
algebra is the space $\Lambda^1(M)/ dC^{\infty}(M)$ of non-exact
one-forms on $M$ which can be identified with ${\cal X}_{div}(M)$
via $L^2-$inner product \cite{mwe83}.
In particular, identifying the velocity field $v$ with the one-form
$v^{\flat}$ via the isomorphism defined by the metric of $M$,
the dynamical formulation of an ideal fluid as geodesic motion
on $Diff_{vol}(M)$ can be reduced to the Lie-Poisson dynamics
\begin{equation}
     {\partial v^{\flat} \over \partial t} =
    - ad^*_{\delta H / \delta v^{\flat}}(v^{\flat})        \label{lpeq}
\end{equation}
on ${\cal X}_{div}(M)$ \cite{arn66}-\cite{MR}.
The physical problem of describing the reduced phase space of the Eulerian
equations (\ref{lpeq}) of hydrodynamics, or the equivalent mathematical
problem of the description of orbits of coadjoint action for the group of
volume preserving diffeomorphisms involves intersections of
infinitely many Casimir functionals of the Lie-Poisson structure which,
by definition, satisfy
\begin{equation}
 \{ H,C \}_{+LP} (v) = - \int_M \; v \cdot [ {\delta H \over \delta v} ,
         {\delta C \over \delta v} ] \; \mu_M  = 0        \label{liep}
\end{equation}
for all functionals $H$ on the reduced phase space. That means, they are
invariants of any dynamics described by Eq.(\ref{lpeq}) on coadjoint
orbits and hence characterize the reduced phase space rather
than the reduced dynamical equations
\cite{ARN},\cite{khc89},\cite{ark92}.

For (+)Lie-Poisson structure described by the bracket in Eq.(\ref{liep}),
the Casimirs are left invariant functions
on orbits associated with the right action of $Diff_{vol}(M)$.
In fluid mechanical context, the right action corresponds to
the particle relabelling symmetries while the left action generates
the motion \cite{djr86},\cite{MR}.
This group theoretical description of motion is essentially
independent of the dimension of the flow space $M$.
In spite of this fact, qualitatively different results were obtained
for Casimirs of even and odd dimensional flows
\cite{khc89}-\cite{ser84}.
\begin{theorem}        \label{prop1}
Let $M$ be a Riemannian manifold of dimension $m$ with a volume $\mu_M$.
The Eulerian equations (\ref{lpeq}) have infinitely
many generalized enstrophy type integrals
\begin{equation}
 I_{\Phi}(v) = \int_M \; \Phi((d_M v^{\flat})^{n} / \mu_M) \mu_M \label{iodd}
\end{equation}
if $m=2n$ and, for $m=2n+1$ there exist at least one generalized
helicity invariant
\begin{equation}
 I(v) = \int_M \; v^{\flat} \wedge (d_M v^{\flat})^{n}  \label{ieven}
\end{equation}
where $d_M$ is the exterior derivative on $M$ and $v^{\flat}$ is the
one-form obtained by lowering the indices of $v$ by the metric on $M$.
\end{theorem}
Apart from incompressible fluids, the physical framework of this
result has been shown in Ref. \cite{khc89},
with slight modifications, to include the
equations of superconductivity \cite{FEYN},\cite{hok83},
barotropic fluids \cite{hmr83} and ideal
magnetohydrodynamics \cite{vid78},\cite{hok83a}.

It has been concluded in Ref. \cite{khc89} that there
might be the possibility of connecting the integral invariants
(\ref{iodd}) and (\ref{ieven})
to symplectic properties of the space of trajectories of the
velocity field.
In fact, for two-dimensional flows, this connection is well-understood
in the framework of the natural symplectic structure of $M$ defined
by its volume two-form. In this case, one can represent the generators
of the symmetry algebra ${\cal X}_{div}(M)$ by Hamiltonian vector fields and
hence it can be identified with the space of nonconstant functions
on $M$ endowed with the canonical Poisson bracket (see section (\ref{ssii}
and Refs. \cite{mwe83},\cite{mor98}).
It then follows that for each infinitesimal symmetry of the velocity
field one can associate a continuous family of Casimirs of the form
of (\ref{iodd}) depending on its Hamiltonian (or stream) function
\cite{vcs90}.

In this work, relying on a symplectic set-up analogous to the one
in two-dimensions, we shall construct infinite families
of helicity integrals for three dimensional flows.
The geometric framework to be employed will also enable us to
investigate the kinematical properties of invariants in the context of
Jacobi structures which includes Poisson, symplectic, conformally
symplectic and contact structures as particular cases.

\subsection{Content of the work}

In Ref. \cite{hg97}, we introduced, for incompressible flows on
a three-dimensional region $M$ of Euclidean space, a symplectic
structure on $R \times M$. Using the automorphism algebra of this structure
we obtained, in Ref. \cite{hg98}, the generators
of volume preserving diffeomorphisms on $M$ and showed that they can,
as in two dimensions, be represented by Hamiltonian vector fields.
This enabled us to express the Lie-Poisson bracket through the
Poisson bracket of invariant functions on $M$.
In this work, we shall utilize these results, which will be
summarized in the next section, to construct infinite families
of helicity invariants and to obtain a kinematical interpretation
of them in the framework of particle relabelling symmetries.

In section (3), associated to each infinitesimal symmetry we shall introduce
invariant two-forms $\Theta_k$ which are closed and degenerate.
This will enable us to express conservation laws globally as the
closure of the three-forms $\theta_k \wedge \Theta_l$ where
$\theta_k$ are potential one-forms satisfying $d \theta_k = \Theta_k$.
The densities which are conserved at each point of trajectories
will be called Lagrangian. By an Eulerian conservation law we shall
mean a divergence expression in which the integral over fluid domain
of a density is conserved \cite{pad98}.
We shall show that the type of conservation laws is determined
by different classes of one-forms $\theta_k$ characterized by
their orientation and invariance properties with respect to
the flow of the velocity field.

In section (4), we shall give a characterization
of the connection between Lagrangian and Eulerian conservations of
helicity in the framework of Jacobi structures on $R \times M$.
We shall first extent the two-forms $\Theta_k$ to symplectic
forms on $R \times M$ without altering their invariance properties.
We shall then establish the correspondences
\begin{equation}
\begin{array}{ccccc}
\begin{array}{c} relative  \\ invariants \end{array} & \leftrightarrow  &
\begin{array}{c} Eulerian  \\ conservation \\ laws \end{array}
& \leftrightarrow  &    \begin{array}{c}  conformally \\ symplectic \\
structures \end{array}     \end{array}   \label{rel1}
\end{equation}
characterized by invariance with respect to flow.
Each family is parametrized by functions on $R \times M$ not
in the kernel of $\partial_t+v$. Moreover, we shall show that
they are conformally
equivalent to the corresponding families in the relations
\begin{equation}
\begin{array}{ccccc}
\begin{array}{c} absolute  \\ invariants \end{array} & \leftrightarrow  &
\begin{array}{c} Lagrangian  \\ conservation \\ laws \end{array}
& \leftrightarrow  &    \begin{array}{c}  symplectic \\
structures \end{array}  \;\;\; .   \end{array}   \label{rel2}
\end{equation}
We shall conclude that the kinematical interpretation of invariants
of coadjoint orbits are connected with the conformal properties
of the space of trajectories, and these can be understood better
in the framework of local Lie algebraic structures on the function
spaces over $R \times M$ rather than with the geometry of
$Diff_{vol}(M)$.

\section{Kinematical symmetries}

The motion of an incompressible fluid in Lagrangian coordinates
can be described as geodesic motion on the group $Diff_{vol}(M)$
of volume preserving diffeomorphisms of $M$ via left action
by evaluation. The right action of the group generates the
particle relabelling symmetries.
A divergence-free frozen in vector field can be used to cast
the suspended velocity field on $R \times M$ into Hamiltonian
form. Under certain conditions automorphisms of the symplectic
structure can be identified with the infinitesimal time-dependent
symmetries on $R \times M$ of the velocity field.
The velocity field itself separates the infinitesimal symmetries
into generators of reparametrizations and diffeomorphisms of $M$.
All these generators can be realized as Hamiltonian vector fields.
These results will be summarized from Refs. \cite{hg97} and \cite{hg98}.

\subsection{Kinematical description and symplectic structure}

\label{mot}

Let the open set $M_0 \subset R^{3}$ be the domain occupied initially
by an incompressible fluid and 
$x(t=0)=x_{0} \in M_0$ be the initial position, i.e., a Lagrangian label.
For a fixed initial position $x_{0}$,
the Eulerian coordinates $x(t) = g_{t}(x_{0})$ define 
a smooth curve in $R^{3}$
describing the evolution of fluid particles. 
For each time $t \in I \subset R$, the volume preserving embedding 
$g_{t}:M_0 \to g_t(M_0)=M \subset R^{3}$ describes a
configuration of fluid. A flow is then a curve $t \mapsto g_{t}$ 
in the group $Diff_{vol}(M)$ of volume preserving diffeomorphisms.
The time-dependent Eulerian (spatial) velocity field $v_{t}$
that generates $g_{t}$ is defined by
\begin{equation}
    {dx \over dt}={dg_{t}(x_{0}) \over dt}
            =(v_{t} \circ g_{t})(x_{0})=v(t,x)    \label{vel}
\end{equation}
where $v_{t} \circ g_{t}$ is the corresponding Lagrangian (material)
velocity field \cite{arn66},\cite{djr86},\cite{MR}.
Since $g_{t}$ is volume preserving, $v_{t}(x)$ is a divergence-free vector
field over $M$ and Eq.(\ref{vel}) is a non-autonomous dynamical system
associated with it.
The Lagrangian description of fluid motion is the description
by trajectories
\cite{djr86},\cite{swh68}-\cite{frv90},
that is, by solutions
of non-autonomous ordinary differential equations (\ref{vel})
or, equivalently, by solutions of the autonomous system represented 
by the suspended velocity field
\begin{equation}
  \partial_{t} + v(t,x)   \;,\;\;\;\;\; v={\bf v} \cdot \nabla \label{tvel}
\end{equation}
on the time-extended space $I \times M$.

The velocity field is
right invariant. Hence, the generators of the right action which
form the infinite dimensional left Lie algebra of $Diff_{vol}(M)$
are infinitesimal particle relabelling symmetries.
The dynamical formulation on $T^*Diff_{vol}(M)$ when reduced by
these symmetries results in the (+)Lie-Poisson structure on 
${\cal X}_{div}^*(M)$. The Eulerian dynamics on the coadjoint
orbits is determined by a right-invariant functional on $I \times M$.
The Eulerian dynamical equations can be used to
construct a formal symplectic structure for (\ref{tvel})
on a time-extended domain $I \times M$ \cite{hg97},\cite{hg98}.
\begin{theorem}           \label{pro2}
In the Eulerian description of motion of an incompressible fluid
in 3D let the dynamics of the velocity field ${\bf v}$ be governed by
\begin{equation}
  {\partial {\bf v} \over \partial t} +
  {\bf v} \cdot \nabla {\bf v}= {\bf F}   \label{geuler}
\end{equation}
and assume that the divergence-free vector field $B$ and the function
$\varphi$ satisfy
\begin{equation}
    {\partial B \over \partial t} + [ v, B]=0  \;,\;\;\;
    {\partial \varphi \over \partial t} +
      v ( \varphi) =0     \label{beq}
\end{equation}
which are the frozen-field equations. Then, $\partial_{t}+v$
is a Hamiltonian vector field
with the symplectic two-form
\begin{equation}
  \Omega = \omega + \sigma \wedge dt \;,\;\;\;
  \sigma =  i(v)(\omega) - d_M \varphi   \;,\;\;\;
  \omega = {\bf B} \cdot (d{\bf x} \wedge d{\bf x})    \label{symp2}
\end{equation}
and the Hamiltonian function $\varphi$. Here, $i(v)( \cdot)$ is the
interior product with the vector field $v$. The invariant density
in the symplectic volume
\begin{equation}
  \mu \equiv {1 \over 2}\Omega \wedge \Omega = 
      \rho_{\varphi} dt \wedge dx \wedge dy \wedge dz   \label{vol}
\end{equation}
is given by $\rho_{\varphi} \equiv  B( \varphi)$

\end{theorem}
Following Ref. \cite{hg98}
we shall show that the symplectic set-up of proposition (\ref{pro2}) for
three dimensional flows is an appropriate modifications of natural
geometric tools of two dimensional flows in the sense that it
enables us to construct the generators of volume preserving
diffeomorphisms and to represent them by Hamiltonian vector fields on $M$.

\subsection{Reparametrization and particle relabelling symmetries}

A time-dependent vector field $U = \xi \partial_{t}+u$ on $I \times M$
is an infinitesimal geometric symmetry of the Lagrangian motion on $M$
described by $v$ if the criterion
\begin{equation}
  [ \partial_{t}+v, \xi \partial_{t}+u ] = 
                (\xi_{,t}+v(\xi))(\partial_{t}+v)      \label{symm}
\end{equation}
is satisfied. These are the most general symmetries of the system
(\ref{vel}) of first order ordinary differential equations \cite{OLV}.
Starting with the Hamiltonian vector fields
\begin{equation}
   u_0 \equiv - \rho^{-1}_{\varphi}B\;,\;\;\;
   U_1 = -u_0(h_1) (\partial_{t} +v) + {dh_1 \over dt} u_0 
    - \rho^{-1}_{\varphi}\nabla \varphi \times \nabla h_1
    \cdot \nabla    \label{fluid}
\end{equation}
associated with the symplectic two-form $\Omega$ and the functions $t$
and $h_1$, where $h_1$ is arbitrary, one can generate infinitely
many infinitesimal automorphisms of $\Omega$. These are vector fields
satisfying ${\cal L}_{U_k}(\Omega)=0$ where ${\cal L}_{U_k}(\cdot)$
is the Lie derivative. The automorphism algebra of
$\Omega$ can be identified with infinitesimal symmetries of $v$
if $dh_1/dt=f(\varphi)$ for some function $f$.
In this case, the vector fields
\begin{equation}
  u_0 \;,\;\;\; U_1\;,\;\;\;
  U_k  \equiv  ({\cal L}_{U_1})^{k-1}(u_0) \;,\;\; k \geq 2
\label{infsym}           \end{equation}
generate an infinite hierarchy of time-dependent infinitesimal
Hamiltonian symmetries of the velocity field $v$.

In order to relate these symmetries to the volume preserving
diffeomorphisms of $M$ it will be appropriate to adopt a coordinate
independent definition of the dynamical system (\ref{vel}) associated
with $v$ because the velocity field is defined only implicitly by some
non-linear Eulerian dynamical equations.
This can be achieved by the interpretation of the system (\ref{vel})
as an algebraic variety $\{ \dot{x}-v(t,x) = 0 \} \subset
J^1_{t,x,1,\dot{x}}(I \times M)$ of the first jet space over $(t,x)$.
This can be embedded into $I \times T_xM$ and thus, Eqs.(\ref{vel})
define a section of the first jet bundle over $I \times M$
represented by $\partial_t+v$. As for any such section, this
induces the unique connection $\Gamma \equiv dt \otimes (\partial_t+v)$ on
$I \times M \to I$ \cite{hg98},\cite{CM}-\cite{SAUN}.
The connection $\Gamma$ on $I \times M$ dictated by the velocity
field $v$ splits the vector fields $U_k$ of the form
$\xi_k \partial_t + u_k$ into horizontal and vertical generators
\begin{equation}
      U^h_k= \xi_k (\partial_t+v) \;,\;\;\;
      U^v_k =u_k - \xi_k v
\end{equation}
of reparametrization symmetries which are gauge transformations and
of diffeomorphisms on $M$, respectively.
Here, $\xi_k$'s are conserved functions of the velocity field
and hence we can identify the algebra of reparametrization symmetries
with the kernel of $\partial_t+v$ in $C^{\infty}(I \times M)$.
$U^v_k$'s are divergence-free vector fields on $M$ with respect to
the time-dependent volume
\begin{equation}
   \mu_M \equiv i(\partial_t)(\mu)= \rho_{\varphi} dx \wedge dy \wedge dz
= - \sigma \wedge \omega = d_M \varphi \wedge \omega
\label{mum}       \end{equation}
on $M$ induced from the symplectic volume
if and only if $h_1$ is conserved under the flow of $\partial_t+v$.
This greatly simplifies the form of vector fields (\ref{infsym}) to
\begin{equation}
   U_k = \xi_k (\partial_t+v) + W_k \;,\;\;\;\;
      u_k \equiv \xi_k v + W_k      \label{uvw}  
\end{equation}
where the left-invariant vector fields on $M$
\begin{equation}
 W_1 \equiv \rho^{-1}_{\varphi}  \nabla h_1
 \times \nabla \varphi \cdot \nabla   \;,\;\;\;
  W_k  \equiv  ({\cal L}_{W_1})^{k-1}(u_0) \;,\;\; k \geq 2    \label{wk}
\end{equation}
are $\mu_M$-divergence free. Introducing the time-dependent functions
\begin{equation}
  \xi_1 \equiv - u_0(h_1) \;,\;\;\;
  \xi_k \equiv - u_0(h_k) \;,\;\;\;
  h_k \equiv  (W_1)^{k-2}(\xi_1) \;,\;\;\; k \geq 2
\label{hwk}        \end{equation}
which are in the form of potential vorticities \cite{sal88} we have
\begin{equation}
  W_k = \rho^{-1}_{\varphi}  \nabla \varphi
   \times \nabla h_k  \cdot \nabla
    \;,\;\;\; k \geq 2           \label{vert}
\end{equation}
and these satisfy the Lie bracket relations
\begin{equation}
  [ W_k, W_l ] =  \rho^{-1}_{\varphi} \nabla \varphi \times
  \nabla h_{lk} \cdot \nabla  \equiv - W_{kl}   \label{wkl}
\end{equation}
of the left Lie algebra of $Diff_{vol}(M)$. The invariant functions
\begin{equation}
    h_{lk} \equiv \rho^{-1}_{\varphi}  \nabla \varphi \cdot
          \nabla h_l \times \nabla h_k     \label{hkl}
\end{equation}
and hence $W_{kl}$ are antisymmetric in their indices.
For each element of the hierarchies $U_k, U_{kl}$ this
process can be continued to find time-dependent, $\mu_M$-divergence-free
vector fields $u_0,W_k,W_{kl},...$ on $M$ which commute with the
suspension $\partial_t+v$.

\subsection{Hamiltonian structures of symmetries}

The vector fields $U_k$ are Hamiltonian with the symplectic two-form
(\ref{symp2}) and the Hamiltonian functions $h_k$. For the generators
$W_k$ of volume preserving diffeomorphisms we have, from Ref. \cite{hg98}
\begin{theorem}
$W_k$'s are manifestly Hamiltonian with the Nambu-Poisson type bracket
\begin{equation}
  \{ f,g \}_{\varphi} = \rho^{-1}_{\varphi} \nabla \varphi
  \cdot \nabla f \times \nabla g  \;,           \label{brm}
\end{equation}
characterized by the function $\varphi$, and with the Hamiltonian
functions $h_k$. The closed two-forms
\begin{equation}
 - \omega_k \equiv  i(W_k)(\mu_M) = d_M \varphi \wedge i(W_k)(\omega)
 \label{hamw}      \end{equation}
on $M$ can be identified with the left-invariant elements of
${\cal X}^*_{div}(M)$.
\end{theorem}
The first equality in Eqs.(\ref{hamw}) is the invariant definition
of the curl vector \cite{khc89},\cite{ark92} and it implies the
Clebsch representations
\begin{equation}
   \rho_{\varphi} {\bf W}_k = \nabla \times \varphi \nabla h_k
         = - \nabla \times h_k \nabla \varphi             \label{cle}
\end{equation}
of $W_k$'s. Since
\begin{equation}
   i(W_k)(\omega) = {\bf B} \times {\bf W}_k \cdot d {\bf x}
    =  d_Mh_k + \xi_k d_M \varphi
\end{equation}
we also conclude from Eqs.(\ref{hamw}) that
the two-forms $\omega_k$ are exact $\omega_k =  d_M \gamma_k$
for one-forms $\gamma_k \in \Lambda^1(M) / d_M C^{\infty}(M)$ defined
up to differential of functions on $M$. Conversely,
since the map $d_M : \Lambda^1(M) / d_M C^{\infty}(M) \to
Image(d_M) \subset \Lambda^2(M) $ does not depend on the representatives
we have the identifications
\begin{equation}
   [ \gamma_k ] \leftrightarrow \omega_k \leftrightarrow W_k
\end{equation}
between equivalence classes of one-forms modulo exact one-forms,
closed two-forms \cite{mwe83},\cite{khc89},\cite{MR}
and the generators $W_k$ of volume preserving diffeomorphisms. 
The Lie bracket algebra of left-invariant vector fields $W_k$
is isomorphic, via
\begin{equation}
  W_{ \{ h_k,h_l \} } = -W_{kl} = [ W_k,W_l ]   \;,  \label{lai}
\end{equation}
to the Poisson bracket algebra (\ref{brm}) of generalized potential
vorticities on the flow space $M$.
Analogous to the canonical Poisson bracket for two dimensional
flows, the Hamiltonian structure on $M$ of the vector fields
$W_k$ can be used to write the (+)Lie-Poisson bracket in three dimensions
in terms of the Poisson bracket (\ref{brm}) on $M$.

\subsection{Nilpotent generators}

The Poisson bracket (\ref{brm}) is
degenerate and possesses a Casimir function on $I \times M$.
If this is one of the functions $h_k$ for some $k >1$, then we have,
by comparing Eqs.(\ref{brm}) and (\ref{vert}), that $W_k=0$.
It follows from Eqs.(\ref{wk}) that
\begin{equation}
   {\cal L}_{W_1}(W_l) =({\cal L}_{W_1})^l(u_0) =
   ({\cal L}_{W_1})^{l-k+1}(W_k) \equiv 0 \;,\;\;\; \forall \; l \geq k \; .
\end{equation}
This, together with Eqs.(\ref{wkl}) and (\ref{hkl}) imply
that $W_1$ is a nilpotent element of the (possibly infinite
dimensional) algebra generated by the finite set $\{ u_0, W_1,...
,W_{k-1} \}$ of vector fields.
Then, by Jacobson-Morozov theorem \cite{VAR},\cite{POST}
there exist vector fields, say $W_0,W_{-1}$, in this finitely
generated algebra satisfying the Lie bracket relations
\begin{equation}
  [ W_1,W_0]=2W_1 \;,\;\;\;  [ W_1,W_{-1}]=W_0 \;,\;\;\;
    [ W_{-1},W_0]=-2W_{-1}
\end{equation}
of the $sl(2,R)$ algebra. Even though the Casimirs of the
bracket (\ref{brm}) gives zero functional on the orbits, the geometric
structures arising from this case, that is, from the nilpotency of $W_1$
is non-trivial and results in Godbillon-Vey type invariants
\cite{ark92},\cite{tab90}-\cite{hag98}.
We refer to Ref. \cite{hag98} for an investigation of this case which
requires a separate treatment, its relation with the symplectic
structure $\Omega$ as well as physically relevant applications.
To this end, we shall solely assume that $W_1$ is not a nilpotent
element of ${\cal X}_{div}(M)$. In other words, the Casimir of
(\ref{brm}) is different from the invariant functions $h_k$ of
potential vorticity type.

\section{Helicity conservations}

\label{invariant}

We shall construct invariant differential forms of the velocity
field associated with the infinitesimal symmetries.
We shall then express the conservation laws as closure of three-forms
obtained from various combinations of invariants. The resulting
divergence expressions imply that the integral over the flow domain
of a density is conserved. As in Ref. \cite{pad98},
these will be called Eulerian
conservation laws. Under certain conditions the divergence
expression reduces to the vanishing of the time derivative
of the density itself. That means, the density is conserved
at each point of the flow domain. This will be called a Lagrangian
conservation law. We shall show that the distinction between
types of conservation laws is kinematical and can be characterized
by gauge transformations on invariant forms.

\subsection{Symplectic structure and integral invariants}

\label{ssii}

We shall discuss and compare the relations between symplectic
structures, Eulerian equations, infinitesimal symmetries and
integral invariants for two and three dimensional flows. 
The ideas to be employed in the rest of this section will
rely on these observations. The construction of infinite
families of helicity integrals will be motivated by
proposition (\ref{prop4}) connecting the symplectic two-form
$\Omega$ to the integral invariant $I(v)$.

The time-dependent, divergence-free velocity field $v$ on
a two-dimensional domain $M$ with coordinates $(x,y)$, and its
curl vector field $w$ perpendicular to $M$ can be expressed by
means of a function $\psi = \psi (t,x,y)$ as
\begin{equation}
  v = { \partial \psi \over \partial x}
      { \partial \over \partial y}
     - { \partial \psi \over \partial y}
      { \partial  \over \partial x}   \;,\;\;\;
      w = \phi { \partial \over \partial z} \;,\;\;\;
      \phi \equiv \nabla^2 \psi          \label{2dv}
\end{equation}
and they satisfy the frozen-field equation for $w$
\begin{equation}
    { \partial \phi \over \partial t}
         + \{ \phi , \psi \}_{can}  =0          \label{ffw}
\end{equation}
where $\{ \;,\; \}_{can}$ is the canonical bracket on the
two-dimensional domain $M$.
The Hamilton's equations (\ref{ffw}) are equivalent to the Euler equations
of ideal fluid in two dimensions and by the Lie algebra isomorphism
(\ref{lai}) to the condition for $w$ to be an infinitesimal time-dependent
symmetry of $v$ \cite{she92},\cite{MR}.

We observe that the formal restriction of the symplectic two-form
(\ref{symp2}) to the vector fields (\ref{2dv}) manifests its interplay
with the two dimensional Eulerian dynamical equations. Namely, we find that
the degenerate two-form
\begin{equation}
   \Omega = - (d \varphi + \phi d \psi ) \wedge dt
             + \phi dx \wedge dy
\end{equation}
is closed whenever Eq.(\ref{ffw}) holds. Moreover, the suspended
velocity field $\partial_t+v$ in three dimensions
is Hamiltonian provided the Hamiltonian function $\varphi$
satisfies the same equation.

To reveal the connection between the symplectic structure
$\Omega$ and the helicity invariant $I(v)$ of proposition (\ref{prop1})
we shall consider, in the notation of proposition (\ref{prop1})
or of Refs. \cite{MR}-\cite{ark92},
the Lie-Poisson equations (\ref{lpeq}) for the kinetic energy
functional, that is, the Euler equations of ideal fluids.
We recall that
a differential $p-$form $\alpha$ is said to be a relative
invariant for a vector field $V$ if there exist a $p-1-$form
$\beta$ such that
\begin{equation}
     {\cal L}_{V}(\alpha) = d \beta \;.
\end{equation}
If $\beta =0$, $\alpha$ is said to be an absolute invariant
\cite{OLV},\cite{AMR},\cite{LM}.
\begin{theorem}        \label{prop4}
On a three-dimensional Riemannian manifold $M$ the Euler equations 
of ideal fluids for a divergence-free vector field $v$ tangent to
the boundary of $M$ are
\begin{equation}
  {\partial v \over \partial t} + \nabla_{v} v = - grad(p)   \label{rel}
\end{equation}
where $p$ is the pressure and $grad$ is taken with respect to the
metric on $M$. Define the exact two-form $\omega$ by
\begin{equation}
    \omega= d_M v^{\flat} \equiv i(w)(\mu_M)         \label{vort}
\end{equation}
where the second equality is the invariant definition of the curl vector
or the vorticity $w$. Then, $\omega$ is an absolute invariant of
$\partial_t+v$, or equivalently,
$w$ is an infinitesimal symmetry of $v$, that is, a frozen-in field.
The symplectic two-form is exact
\begin{equation}
   \Omega = - d \theta  \;,\;\;\;\;
    \theta = (\varphi + p + {1 \over 2} v^2) dt - v^{\flat}
\end{equation}
and it is an extention to $I \times M$ of $\omega$ on $M$ via
Euler equations (\ref{rel}).
$\theta$ and $\theta \wedge \Omega$ are relative invariants.
The integrand in $I(v)$ of proposition (\ref{prop1})
is the scalar density in the three-form
$\theta \wedge \Omega$ and is associated with the infinitesimal
symmetry $w$ of the velocity field. The identity
\begin{equation}
   d(\theta \wedge \Omega) + \Omega  \wedge \Omega \equiv 0  \;.
\end{equation}
is an expression for the (Eulerian) conservation law of helicity
in divergence form.
\end{theorem}
{\bf Proof:}
Using the identity $(\nabla_v v)^{\flat} = {\cal L}_{v}(v^{\flat})
- d_M v^2/2$ in Eq.(\ref{rel}) and taking the derivative $d_M$
of resulting equation we obtain $\omega_{,t}+{\cal L}_{v}(\omega)=0$
\cite{MR}. So, $\omega$ is an absolute invariant for $\partial_t+v$.
In terms of $\omega = i(w)(\mu_M)$ this gives
\begin{eqnarray}
 0&=& (i(w)(\mu_M))_{,t}+ {\cal L}_v(i(w)(\mu_M))      \\
  &=& i(w_{,t})(\mu_M) + i(w)((\mu_M)_{,t}) + i(w)({\cal L}_v(\mu_M))
          + i([v,w])(\mu_M)        \\
  &=& i(w_{,t} + [v,w])(\mu_M) + i(w)((\mu_M)_{,t} + {\cal L}_v(\mu_M))
\label{symww}   \end{eqnarray}
where we used the identity ${\cal L}_v \circ i(w) - i(w) \circ
{\cal L}_v = i([v,w])$ \cite{MR} in obtaining the second equation. Since
$\mu_M$ is invariant, the second term in Eq.(\ref{symww}) vanishes
and the fact that it defines a volume implies $w_{,t} + [v,w]=0$.
Thus, $w$ is an infinitesimal symmetry of $v$.

Solving $v^{\flat}_{,t}$ from the Euler equations in the derivative
of $\theta$ one obtains the symplectic two-form $\Omega$.
Equivalently, it can also be obtained from $\omega=d_M v^{\flat}$
by replacing $d_M$ with $d$, solving the time derivative of
the velocity field from the Euler equations (\ref{rel}) and adding
the one-form $\varphi dt$ with $\varphi$ being any conserved function
of $v$. The Lie derivatives
\begin{equation}
   {\cal L}_{\partial_t+v}(\theta) = d \chi \;,\;\;\;
   {\cal L}_{\partial_t+v}(\theta \wedge \Omega)=d(\chi \Omega)
  \;,\;\;\; \chi \equiv \varphi +p -{1 \over 2} v^2    \label{echi}
\end{equation}
express the relative invariances of $\theta$ and $\theta \wedge \Omega$.
For the last conclusion, we compute
\begin{eqnarray}
  0&=&d(\theta \wedge \Omega) + \Omega  \wedge \Omega    \\
   &=&d[ v^{\flat} \wedge d_M v^{\flat} + [ (\varphi + p + {1 \over 2} v^2)
         d_M v^{\flat} - v^{\flat} \wedge
      (d_M \varphi - i(v)(\omega)) ] \wedge dt ]    \nonumber  \\
  & & \;\;\;\;\;\;\;\;\;\;\;\;\;\;
    +2 d_M v^{\flat} \wedge (d_M \varphi - i(v)(\omega)) \wedge dt  \\
   &=& [ (v^{\flat} \wedge d_M v^{\flat})_{,t} +
   d_M (( p + {1 \over 2} v^2) d_M v^{\flat}
   + i(v)(\omega) \wedge  d_M v^{\flat}  )    ] \wedge dt   \label{orhe}
\end{eqnarray}
which is the divergence expression for the local form of the conservation
law for the total helicity $I(v)$.   $\bullet$

Note that the helicity flux in Eq.(\ref{orhe}) is independent of
the function $\varphi$ which we have introduced by hand to make the
symplectic form non-degenerate.
The function $\chi$ in the invariance expressions (\ref{echi}) is
related, in Ref. \cite{pam96}, to the invariance under the particle
relabelling symmetries of the Lagrangian density of the
variational formulation of Eqs.(\ref{rel}) in which $\varphi$
corresponds to the sum of the potential energies of the fluid
\cite{swh68},\cite{sal88}.

\subsection{Invariant differential forms}

Proposition (\ref{prop4}) explains the connection between
conservation law for the helicity integral $I(v)$ and the
curl $w$ of $v$ viewed as an infinitesimal symmetry. It,
moreover, gives a recipe to construct the conserved density
in $I(v)$ starting from $w$. We shall now apply this to the
generators $W_k$ of particle relabelling symmetries
and obtain infinitely many integrals of the form of $I(v)$.
Our presentation of invariant forms will be three-fold
(c.f. Eqs.(\ref{bigto}-(\ref{bigt})). The abstract coordinate independent
form of them (c.f. Eqs.(\ref{bigto})) will serve for generalization
to and for computation (as in proposition (\ref{prop4})) on any
Riemannian manifold $M$.
The Clebsch representation of them (c.f. Eqs.(\ref{bigtot}))
will follow from our earlier results presented in section (2).
The coordinate expressions in the notations of three-dimensional
vector calculus (c.f. Eqs.(\ref{bigt})) will be used to make
the results more excessible as well as to justify the invariant
formulation.
\begin{theorem}
The exact, degenerate two-forms
\begin{eqnarray}
 \Theta_k &=& \omega_k + i(v)(\omega_k) \wedge dt   \label{bigto}  \\
        &=& d \varphi \wedge d h_k   \label{bigtot}       \\
     &=& - \rho_{\varphi} [ {\bf W}_k \cdot d{\bf x} \wedge d{\bf x}
     +( {\bf W}_k \times {\bf v}) \cdot d{\bf x} \wedge dt ]
\label{bigt}       \end{eqnarray}
are absolute invariants of the velocity field. They are the
extentions to the space $I \times M$ of $\omega_k$'s
and can be obtained by replacing $d_M$ in Eq.(\ref{hamw}) by
$d=d_M + dt \wedge \partial_t$.
\end{theorem}
{\bf Proof:} The degeneracy
\begin{equation}
   \Theta_k   \wedge \Theta_l = 0 \;,\;\;\;  \forall \; k,l  \label{deg}
\end{equation}
can be seen by direct computation.
The closure and absolute invariance of $\omega_k$'s imply
via Eq.(\ref{bigto}) the closure of $\Theta_k$'s.
For Eq.(\ref{bigt}) the closure of $\Theta_k$ follows from
the conservations of
$\rho_{\varphi}$, $\nabla \cdot {\bf v}=0, \; \nabla \cdot
(\rho_{\varphi} {\bf W}_k) =0$ and the left invariance of $W_k$'s.
The absolute invariance follows from the closure of $\Theta_k$ and
that it annihilates the extended velocity field.
Employing the Poincar\'{e} lemma, we introduce potential
one-forms $\theta_k$
\begin{equation}
 \Theta_k =  d \theta_k \;,\;\;\;
  - \theta_k = \psi_k dt +  A_k \;\;\;\;
     A_k \equiv  {\bf A}_k \cdot d{\bf x}           \label{steta}
\end{equation}
where $\psi_k$ and the representative $A_k$ of the one-form $\gamma_k$
satisfying $d_M \gamma_k = \omega_k$ are defined by the equations
\begin{equation}
     \omega_k = d_M A_k \;,\;\;\;\;
     A_{k,t} - i(v)(\omega_k) = d_M \psi_k      \label{psis}
\end{equation}
or equivalently,
\begin{equation}
   \rho_{\varphi} {\bf W}_k =  \nabla
   \times {\bf A}_k    \;,\;\;\;
  {\partial {\bf A}_k \over \partial t} +
  \rho_{\varphi} {\bf W}_k \times {\bf v} =
  \nabla \psi_k            \label{psi}
\end{equation}
the first of which can be regarded as to define the Clebsch potentials
\cite{swh68},\cite{sal88} ${\bf A}_k$ for the vector fields $W_k$.
$\Theta_k$'s can be obtained from $\omega_k=d_M A_k$ by
replacing $d_M$ with $d$ and solving $A_k$ from Eqs.(\ref{steta}).
$\bullet$

The one-forms $\theta_k$ whose derivatives give $\Theta_k$
are defined up to differential of an arbitrary function on $I \times M$.
The invariance properties of $\theta_k$'s are 
characterized by these functions. Since exact forms on
${\cal X}^*_{div}(M)$ result in zero functionals, one-forms
\begin{equation}
 [ \theta_k ] \equiv \{   \theta_k
 + d \lambda_k  \; | \; d \theta_k = \Theta_k ,\;
   \lambda_k  \in C^{\infty}(I \times M)  \}    \label{class}
\end{equation}
constitute an equivalence class on coadjoint orbits. However, they
can be distinguished by the velocity field according to their
behaviour under its flow. To this end, we shall assume that
the one-forms $\theta_k$ as given by Eqs.(\ref{steta}) and (\ref{psis})
are all annihilated by the extended velocity field
\begin{equation}
   i(\partial_t+v)(\theta_k) = - \psi_k - i(v)(A_k) =0
       \;\;\;\;   \forall \; k      \label{kerv}
\end{equation}
which will avoid the proliferation of various exact one-forms
in the foregoing discussions. Having fixed this gauge for
$\omega_k$'s, we compute
\begin{equation}
  {\cal L}_{\partial_t+v}(\theta_k) =
  di(\partial_t+v)(\theta_k) = d \chi_k  \;,\;\;\;\;
  \chi_k \equiv  \lambda_{k,t} + v ( \lambda_k )       \label{chi}
\end{equation}
so that, if $\lambda_k$ is not a conserved function
for the velocity field $\theta_k$'s are only relatively invariant.
Thus, the type of invariance of $\theta_k$'s separates the class
(\ref{class}) into subclasses 
\begin{eqnarray}
 [ \theta_k ]^a \equiv \{  \theta_k
 + d \lambda_k  \; | \; d \theta_k = \Theta_k ,\;
   \lambda_k  \in ker(\partial_t+v) \subset
   C^{\infty}(I \times M)  \}    \label{aclass}        \\ \;
 [ \theta_k ]^r \equiv \{  \theta_k
 + d \lambda_k  \; | \; d \theta_k = \Theta_k ,\;
   \lambda_k  \in C^{\infty}(I \times M) /
      ker(\partial_t+v)  \}     \label{rclass}
\end{eqnarray}
of absolutely and relatively invariant one-forms, respectively.
Here, we identify elements of $ker(\partial_t+v)$ which differ by
an additive term linear in the time variable.
We shall take the representatives of $[ \theta_k ]^a$ as defined
by Eqs.(\ref{steta}) and (\ref{psis}).
We thus have the decomposition
\begin{equation}
 [ \theta_k ] = [ \theta_k ]^a \oplus  [ \theta_k ]^r  \;.    \label{ar}
\end{equation}
of one-forms on $I \times M$ which can alternatively be interpreted
as the splitting of $T^*(I \times M)$ into horizontal and
vertical subspaces by the connection $dt \otimes (\partial_t+v)$.

\subsection{Lagrangian conservation laws}

We shall first construct helicity densities which are conserved
at each point of trajectories of the velocity field.
These Lagrangian conservation laws will be formulated
using a pair of invariant potential one-forms one of which
is in the class $[ \theta_k ]^a$ of absolutely invariant ones.
In the case of Clebsch representations, a proper orientation
of them becomes nesessary.
\begin{theorem}          \label{prop6}
For $k \neq l$ and for $\theta_k \in [ \theta_k ]^a$, the closure
of the three-forms $\theta_k \wedge \Theta_l$ is equivalent to
conservations of helicity densities ${\bf A}_k \cdot {\bf W}_l$
under the flow of the velocity field.
\end{theorem}
{\bf Proof:}
The three-forms are closed identically by the property (\ref{deg})
of the two-forms $\Theta_k$. To obtain the conservation laws we write
\begin{eqnarray}
& &  \theta_k \wedge  \Theta_l 
     = \rho_{\varphi} {\bf A}_k \cdot {\bf W}_l  \;\;
    dx \wedge dy \wedge dz   + \nonumber     \\
 & & \;\;\;\;\;\;\;\;\;\;\;\;\;\;\; 
       \rho_{\varphi}        [ \psi_k {\bf W}_l
    + ({\bf v} \cdot {\bf A}_k) {\bf W}_l
    - ( {\bf W}_l \cdot {\bf A}_k) {\bf v} ]
   \cdot  d{\bf x} \wedge d{\bf x} \wedge dt          \label{thr}  \\
 & & \;\;\;\;\;\;\;\;\;\;\;\;\;\;\; 
= \rho_{\varphi}({\bf A}_k \cdot {\bf W}_l)
   ( dx \wedge dy \wedge dz  - {\bf v} 
   \cdot  d{\bf x} \wedge d{\bf x} \wedge dt  )        \label{thre}
\end{eqnarray}
where we used Eqs.(\ref{steta}),(\ref{psi}), the vector identity
$ {\bf A}_k \times ( {\bf W}_l \times {\bf v}) =
({\bf v} \cdot {\bf A}_k) {\bf W}_l- ( {\bf W}_l \cdot {\bf A}_k) {\bf v}$
and Eq.(\ref{kerv}). Applying $d$ to (\ref{thre}) we get
\begin{equation}
{\partial \over \partial t} ({\bf A}_k \cdot {\bf W}_l)
   + {\bf v} \cdot \nabla ({\bf A}_k \cdot {\bf W}_l) =0
\label{laco}      \end{equation}
which is the expression for a Lagrangian conservation law. $\bullet$

Two particular solutions to Eqs.(\ref{psi}) are given by the one-forms
\begin{equation}
 \theta_k^- = - h_k d \varphi \;,\;\;\; \theta_k^+ =  \varphi d h_k
\label{tpm}    \end{equation}
which are connected with the Clebsch representations (\ref{cle})
of $W_k$'s. The existence of Lagrangian conserved densities of
helicity type depends
on the proper choice of orientation for the potential one-form which
is a topological property. For the above solutions 
\begin{equation}
  \psi_k^- =  h_k \varphi_{,t} \;,\;\;\;
    \psi_k^+ = - \varphi h_{k,t} \;,\;\;\;
  {\bf A}^- =  h_k \nabla \varphi \;,\;\;\;
    {\bf A}^+ = - \varphi \nabla h_k
\end{equation}
of Eqs.(\ref{psi}) which follows from Eqs.(\ref{bigto})
the absolutely invariant three-forms
\begin{eqnarray}
& &  \theta_k^+ \wedge d \epsilon \theta_l^{\epsilon} =
       \theta_k^+ \wedge \Theta_l  =      
       \varphi dh_k \wedge d \varphi \wedge d h_l      \\
 & & \;\;\;\;\;\;\;\;\;\;\;\;\;\;\; 
 = \varphi [ \nabla h_k \times \nabla \varphi \cdot \nabla h_l \;\;
                 dx \wedge dy \wedge dz    \nonumber  \\
 & & \;\;\;\;\;\;\;\;\;\;\;\;\;\;\;\;\;\; 
      \nabla h_l \times (\varphi_{,t} \nabla h_k
      - h_{k,t} \nabla \varphi ) + h_{l,t} \nabla h_k \times
 \nabla \varphi ) \cdot  d{\bf x} \wedge d{\bf x} \wedge dt ] \;\;\;\;\;\;\;
     \label{three}
\end{eqnarray}
are the non-zero products of $\Theta_l$'s with $\theta_k$'s $\forall
\;k \neq l$ whereas $\theta^-_k \wedge \Theta_l \equiv 0 \; \forall \;k,l$.
In Eq.(\ref{three}) the helicity density is recognized to be the
volume density in the three space with coordinates
$(h_k, h_l , \varphi^2/2 )$.

\subsection{Eulerian conservation laws}

The conservation of helicity density at each point of trajectories is,
a consequence of the absolute invariance which follows from the
condition (\ref{kerv}).
The violation of the condition within the cohomology class
of the potential one-forms by a gauge transformation, that is,
by the addition of an exact one-form not in the kernel of
$\partial_t+v$ changes the character of conservation laws.
Since the operators $d$ and ${\cal L}_{\partial_t+v}$ commute
the absolute invariance of two-forms $\Theta_k$ are not affected
by such a transformation.
However, for the corresponding class of three-forms
$\theta_k \wedge d\theta_l$ we have
\begin{equation}
  {\cal L}_{\partial_t+v}( \theta_k \wedge d \theta_l) =
         d ( \chi_k d \theta_l )  
\end{equation}
and hence the distinction by invariance under velocity field
in the class of one-forms
can be carried over to the class of three-forms. This, in turn,
changes completely the character of the Lagrangian conservation laws
\begin{equation}
  d (\theta_k \wedge d \theta_l) =0        \label{hec}
\end{equation}
of proposition (\ref{prop6}). Namely, for each $\chi_k \neq constant$
within a given class of one-form $\theta_k$
one obtains a conservation law of divergence type.
\begin{theorem}    \label{mmm}
(1) Each representative of the class $[\theta_k]$ parametrized
by the function space $C^{\infty}( I \times M)$
gives infinitely many conservation laws of helicity type
expressed as the closure of the three-forms 
$\theta_k \wedge \Theta_l \;, \;\forall \; l \neq k$.

(2) For potential one-forms in $[\theta_k]^a$ the conservation law
is of Lagrangian type (\ref{laco}), while for those in $[\theta_k]^r$
it is an Eulerian conservation law.

(3) For each Lagrangian invariant, there are infinitely many
Eulerian conservation laws parametrized by
$C^{\infty}(I \times M) / ker(\partial_t+v)$ all of
which are equivalent within the class of defining three-forms.

(4) Each Eulerian conservation law associated with
$\theta_k \wedge \Theta_l$ degenerates into the equivalent
Lagrangian one whenever $W_l$ is tangent to the level
surfaces of $\chi_k$.
In this case, the functions $W_l(\lambda_k)$ are
also Lagrangian conserved densities.
\end{theorem}
{\bf Proof:} Representing a one-form $\theta_k$ in the class
(\ref{class}) by
\begin{equation}
  \theta_k = \phi_k dt + {\bf a}_k \cdot d{\bf x} \;,\;\;\;
  \phi_k =  \lambda_{k,t} -  \psi_k
    \;,\;\;\;  {\bf a}_k = \nabla \lambda_k - {\bf A}_k  
\end{equation}  
we compute the three-form
\begin{eqnarray}
& & - \theta_k \wedge d \theta_l
     = \rho_{\varphi} {\bf a}_k \cdot {\bf W}_l  \;\;
    dx \wedge dy \wedge dz   + \nonumber     \\
 & & \;\;\;\;\;\;\;\;\;\;\;\;\;\;\;\;\;\; 
       \rho_{\varphi}        [ \phi_k {\bf W}_l
    + ({\bf v} \cdot {\bf a}_k) {\bf W}_l
    - ( {\bf a}_k \cdot {\bf W}_l) {\bf v} ]
   \cdot  d{\bf x} \wedge d{\bf x} \wedge dt          \label{gthr} 
\end{eqnarray}
to be associated with the helicity type conservation laws.

(1) The closure of (\ref{gthr}) gives
\begin{equation}
{\partial {\cal H}_{kl} \over \partial t} + \nabla \cdot
  [ {\cal H}_{kl} {\bf v} - \rho_{\varphi} \chi_k {\bf W}_l) ] =0
\label{eco}      \end{equation}
for the generalized helicity densities
\begin{equation}
  {\cal H}_{kl} = \rho_{\varphi} {\bf a}_k \cdot {\bf W}_l
         = \rho_{\varphi} ( W_l (\lambda_k)
          - {\bf A}_k \cdot {\bf W}_l )
\label{heden}      \end{equation}
which depend on two discrete and a continuous parameter.
For fixed $k$ and function $\lambda_k$, ${\cal H}_{kl}$ are
indexed by the generators $W_l$ of particle relabelling
symmetries which are infinite in number.
For each fixed pair of discrete parameters $k,l$, the
continuous parameter in ${\cal H}_{kl}$ is the function
$\lambda_k$ and this characterizes the type of conservation
laws.

(2) For $\lambda_k \in ker(\partial_t+v)$, we have, from
Eqs.(\ref{chi}) $\chi_k =0$ and Eq.(\ref{eco}) turns into a
Lagrangian conservation law.

(3) $\forall \; \lambda_k \in ker(\partial_t+v)$ the density
${\cal H}_{kl}$ in Eq.(\ref{eco}) is independent of this
function because the term $W_l (\lambda_k)$
vanishes via commutativity of $\partial_t+v$ and $W_l$.
As $\lambda_k$ takes values in
$C^{\infty}(I \times M) / ker(\partial_t+v)$ 
we obtain infinitely many Eulerian conservation laws with
densities ${\cal H}_{kl}$ having the Lagrangian invariant
${\bf A}_k \cdot {\bf W}_l$ in common and differing only in
the term $W_l (\lambda_k)$. Since we fixed the discrete
parameters, these conservation laws arise from the same
class of three-forms.

(4) Using divergence-free properties of $v$ and $W_l$, the Eulerian
conservation law (\ref{eco}) can be put into the form
\begin{equation}
{\partial {\cal H}_{kl} \over \partial t}  +
   {\bf v} \cdot \nabla {\cal H}_{kl} =
   \rho_{\varphi} W_l(\chi_k)  
\label{ecoo}      \end{equation}
from which the last conclusion follows.
If $W_l(\chi_k)=W_l((\partial_t+v)(\lambda_k)) =0$, we have
\begin{equation}
  { \partial W_l(\lambda_k) \over \partial t} +v(W_l(\lambda_k))=0
\end{equation}
by left-invariance of $W_l$.
$\bullet$

As we remarked earlier for the Clebsch representations of $W_k$'s,
the non-vanishing helicity densities result from the proper
choice of orientation within the class of potential one-forms.
This is a topological property whereas the distinction (\ref{ar})
in the cohomology classes of potential one-forms and
hence in the type of conservation laws is purely kinematical.
Moreover, since the representatives of the class $[\theta_k]$
can not be distinguished on orbits, there is no difference between
Lagrangian and Eulerian conserved densities as invariants of
coadjoint orbits.
The gauge degree of freedom in helicity integrals has been
indicated in \cite{pad98}.
The relation between Lagrangian and Eulerian conservation
laws has not been made clear because the exact one-form
in the representative of $\theta_k$ has been restricted
to be an invariant of the velocity field.

Analogous to the invariants (\ref{ieven}) of even dimensional
flows, the Casimirs
\begin{equation}
  C(\rho,w(\rho)) = \int_M \; \Phi ( \rho, w(\rho)) \; dx \; dy \; dz
\end{equation}
depending on the arbitrary function $\Phi$ has been considered
for three-dimensional motions \cite{pam96}-\cite{ah87}.
However, contrary to (\ref{ieven}) involving Eulerian vorticity
variable, $C(\rho,w(\rho))$ contain Lagrangian information
\cite{she92}.

It easily follows from the symmetry condition (\ref{symm})
that an infinitesimal
symmetry takes a conserved density into another one.
The last conclusion of proposition (\ref{mmm}), on the other hand,
makes it possible to have $W_l(\lambda_k)$ as a conserved function
of the velocity field even if $\lambda_k$ itself is not.
In fact, for negative orientation of the potential one-forms
$\theta_k$ in Eqs.(\ref{tpm}) the conserved densities ${\cal H}_{kl}$
consist only of these functions because we have ${\bf A}_k^- \cdot
{\bf W}_l =0$ in Eq.(\ref{heden}).

\section{Kinematical interpretations}

We shall now seek a characterization of the connection between the
Eulerian conservation laws and Lagrangian invariants
both of which stemm from the one and the same hierarchy of
infinitesimal symmetries, with the kinematical distinction
encoded in the invariance properties in the equivalence classes
of the associated differential
forms with respect to the flow of the velocity field.
We shall find an interpretation of this distinction in the
geometric language of Jacobi structures on $I \times M$
or equivalently, in the Lie algebraic structures on
$C^{\infty}(I \times M)$. More precisely, we shall prove
\begin{theorem}                             \label{ppp}
The connection between Lagrangian invariants and the hierarchies
of Eulerian conservation laws anchored to them is the same as the
conformal equivalence of a Poisson bracket algebra to an infinite
hierarchy of local Lie algebras.
\end{theorem}

\subsection{Extentions of invariant forms}

To establish the result of proposition (\ref{ppp}) we shall consider
further extentions of two-forms $\Theta_k$ to closed two-forms with
maximal rank on $I \times M$, that is, to symplectic forms.
The degeneracy of $\Theta_k$'s can be removed by an additional
term which does not change their closure and invariance
properties. In particular, these local conditions are satisfied
if we demand the extentions of $\Theta_k$'s to be symplectic.
\begin{theorem}
Let $\eta_k$ be a time-dependent, closed and left-invariant
one-form on $M$ which, for $k > 0$, is different from $d_M \varphi$
and $d_M h_k$. Then, $\partial_t+v$ is locally Hamiltonian with the
symplectic two-form
\begin{eqnarray}
 \Omega_k &=& \Theta_k +  \eta_k \wedge dt     \label{omk}  \\
     &=& - [ \rho_{\varphi} {\bf W}_k \cdot d{\bf x} \wedge d{\bf x}
     +( \rho_{\varphi} {\bf W}_k \times {\bf v} - {\bf n}_k)
     \cdot d{\bf x} \wedge dt ]  \;\;.
\label{omek}       \end{eqnarray}
where we let $\eta_k \equiv {\bf n}_k \cdot d {\bf x}$. 
If, moreover, $\partial_t+v$ is globally Hamiltonian, then, there
exists time-dependent function $\varphi_k$ such that
\begin{equation}
     \eta_k = - d_M \varphi_k  \;,\;\;\;
    i(\partial_t+v)( \Omega_k ) =  d \varphi_k   \;,\;\;\; k \geq 0 \;.
\end{equation}
In this case, $\Omega_k$ is exact with the canonical one-form
\begin{equation}
   \tilde{\theta}_k = \varphi_k dt - \theta_k \;,\;\;\;
   \Omega_k = - d \tilde{\theta}_k
\end{equation}
where $\theta_k$ is a representative of the class of one-forms
satisfying $d\theta_k = \Theta_k$. $\theta_k$ and $\tilde{\theta}_k$
have the same invariance properties and the identities
\begin{equation}
  d(\tilde{\theta}_k \wedge \Omega_l)+ \Omega_k \wedge \Omega_l =0 \;.
\end{equation}
give the helicity conservation laws of proposition (\ref{mmm}).
\end{theorem}
{\bf Proof:}
Using Eqs.(\ref{omk}),(\ref{bigto}),(\ref{hamw}) we compute
\begin{eqnarray}
 \Omega_k \wedge \Omega_l &=& d_M \varphi \wedge
      ( \eta_l \wedge d_M h_k + \eta_k \wedge d_M h_l ) \wedge dt   \\
  &=& -( {\bf W}_k \cdot {\bf n}_l + {\bf W}_l \cdot {\bf n}_k ) \; \mu
\end{eqnarray}
where $\mu$ is the symplectic volume defined by $\omega$.
For $k=l$ this gives
\begin{equation}
 \Omega_k \wedge \Omega_k = -2 {\bf W}_k \cdot {\bf n}_k \; \mu
   = -2 \nabla \varphi \times \nabla h_k \cdot {\bf n}_k \; \mu
\end{equation}
and hence the assumptions on $\eta_k$ make $\Omega_k$ to be non-degenerate.
The conditions of closure and absolute invariance of $\Omega_k$ implies
\begin{equation}
     d_M \eta_k =0 \;,\;\;\;
       \eta_{k,t} + d_M i(v)(\eta_k) =0       \label{cinv}
\end{equation}
the first of which makes $\Omega_k$ into a symplectic form.
The second equation is obtained from
\begin{equation}
  {\cal L}_{\partial_t+v}(\Omega_k)
  = d i(\partial_t+v)(\rho_{\varphi} \eta_k \wedge dt) =
    \rho_{\varphi} ( \eta_{k,t} + d_M i(v)(\eta_k)) \wedge dt   
\end{equation}
and expresses the advection of $\eta_k$ by the flow of $v$.
Eqs.(\ref{cinv}) can also be realized as the integrability
conditions for the equations
\begin{equation}
   \eta_k = - d_M \varphi_k \;,\;\;\;
   i(v)(\eta_k) = \varphi_{k,t}             \label{zeta}
\end{equation}
defining the time-dependent function $\varphi_k$ for given $\eta_k$.
By the existence of these functions the suspended velocity field
admits infinitely many symplectic formulations
\begin{equation}
    i(\partial_t+v)( \Omega_k ) =  d \varphi_k   \;,\;\;\; k \geq 0
\label{hamk}      \end{equation}
which, for $k=0, \omega_0=\omega$ and $\varphi_k = \varphi$,
coincide with the one we begin with. Since
\begin{equation}
  {\cal L}_{\partial_t+v}(\varphi_k dt) =
  d \varphi_k + i(\partial_t+v)(d_M \varphi_k \wedge dt) =
  (\varphi_{k,t}+v(\varphi_k)) dt =0
\end{equation}
by Eqs.(\ref{zeta}), we have
\begin{equation}
  {\cal L}_{\partial_t+v}(\tilde{\theta}_k) =
  {\cal L}_{\partial_t+v}(\theta_k) 
\end{equation}
and hence the invariance class of $\tilde{\theta}_k$ is determined by
that of $\theta_k$ in (\ref{ar}). For the extended form of the
helicity conservation laws we compute
\begin{eqnarray}
 0&=& - d(\theta_k \wedge \Theta_l)
  \;=\; - d [ ( \varphi_k dt - \tilde{\theta}_k ) \wedge
            ( \Omega_l - \eta_l \wedge dt ) ]      \\
  &=& d(\tilde{\theta}_k \wedge \Omega_l) + ( \Omega_k - \Theta_k )
       \wedge  \Omega_l + \Omega_k \wedge \eta_l \wedge dt  \\
  &=& d(\tilde{\theta}_k \wedge \Omega_l) + \Omega_k \wedge \Omega_l
\;\;\;\; \bullet     \end{eqnarray}
Thus, the extentions $\Omega_k$ of degenerate, exact two-forms $\Theta_k$ is
induced by a translation of the scalar part of the potential
one-forms $\theta_k$ with a conserved function of the velocity field
\begin{equation}
  \psi_k \mapsto \psi_k + \varphi_k  \; \Rightarrow \;
  \theta_k \mapsto \tilde{\theta}_k  \;,\;\;\;
  \Theta_k = d \theta_k \mapsto \Omega_k = - d \tilde{\theta}_k  \;.
\end{equation}
The canonical one-forms $\tilde{\theta}_k$ are relative invariants of $v$
\begin{equation}
   {\cal L}_{\partial_{t}+v}( \tilde{\theta}_k )= d \varphi_k 
\end{equation}
provided the gauge fixing conditions $\psi_k + {\bf v} \cdot {\bf A}_k=0$
hold and they become absolute invariants whenever $\varphi_k$'s
are constants, that is, on the level surfaces of the Hamiltonian
functions in Eqs.(\ref{hamk}). The three-forms $\tilde{\theta}_k
\wedge \Omega_l$ are also relative invariants
\begin{equation}
   {\cal L}_{\partial_{t}+v}( \tilde{\theta}_k \wedge \Omega_l)=
           d (\varphi_k \Omega_l )
\end{equation}
which when $\varphi_k =constant$ become absolute invariants
because $\Omega_l$ are symplectic.
In this case, the helicity densities are Lagrangian invariants.
This is the particular relation (\ref{rel2}) between invariants, geometric
structures and conservation laws. Below we shall consider
a framework for geometric structures on $I \times M$ more general
than the symplectic one to obtain the relation (\ref{rel1}) between relative
invariances and Eulerian conservation laws.
The corresponding Lie algebraic structure on $C^{\infty}(I \times M)$
is a generalization of the Poisson bracket algebra to the one
which relaxes the Leibniz' rule and this is connected with Jacobi
structures on $I \times M$.

\subsection{Jacobi structures}

A local Lie algebra structure on the space $C^{\infty}(N)$ of smooth
functions on a smooth manifold $N$ is defined by a bilinear mapping
\begin{equation}
  \{ \;,\; \} : C^{\infty}(N) \times C^{\infty}(N) \to C^{\infty}(N) 
\end{equation}
satisfying the conditions of
skew-symmetry $\{ f,g \} =- \{ g,f \} $ and the Jacobi identity
\begin{equation}
  \{ f, \{ g,h \} \} + \{ h, \{ f,g \} \} + \{ g, \{ h,f \} \} =0
\label{fgh}         \end{equation}
for arbitrary $f,g,h \in C^{\infty}(N) $.
The bracket is local in the sense that
\begin{equation}
  support( \{ f,g \} ) \subseteq support(f) \cap support(g)
\end{equation}
and in general, $\{ \;,\; \}$ is not a derivation in its arguments.
The local Lie algebra structure on $C^{\infty}(N)$ is linked
with the Jacobi structure $(\Lambda,E)$ on $N$ through
\begin{equation}
  \{ f,g \} = \Lambda (df \wedge dg) + E(fdg-gdf)
\end{equation}
where the bi-vector field $\Lambda : N \to \Lambda^2(TN) = TN \wedge TN$
and the vector field $E$ on $N$ satisfy the conditions
\begin{equation}
   [ \Lambda,\Lambda ] = 2 E \wedge  \Lambda \;,\;\;\;\;
   [ \Lambda,E ] = 0      \label{jjac}
\end{equation}
imposed by the Jacobi identity (\ref{fgh})
\cite{kir76},\cite{lic78},\cite{LM}.
The coordinate expression of the bracket
\begin{equation}
   [ \Lambda,\Lambda ] = ( \Lambda^{ij}_{,l} \Lambda^{lk} +
   \Lambda^{jk}_{,l} \Lambda^{li} + \Lambda^{ki}_{,l} \Lambda^{lj} )
 \; \partial_i \wedge \partial_j \wedge \partial_k 
\end{equation}
for a bi-vector $\Lambda =  \Lambda^{ij} \partial_i \wedge \partial_j /2$
is familiar from the Jacobi identity for a Poisson tensor and, 
$[ \Lambda,E ] = {\cal L}_{E} (\Lambda)$ is the Lie derivative
\cite{MR},\cite{OLV}.

The Jacobi structure $(\Lambda,E)$ on $N$ includes, as a particular case,
the Poisson structure $\Lambda$ if $E=0$.
When $\Lambda$ is of maximal rank on an even dimensional manifold $N$,
one can define the two-form
$\Omega \equiv \Lambda^{-1}$ and the one-form $\alpha \equiv i(E)(\Omega)$
satisfying the equations
\begin{equation}
   d \Omega = \alpha  \wedge \Omega
   \;,\;\;\;\;\;  d \alpha =0            \label{consym}
\end{equation}
corresponding to Eqs.(\ref{jjac}). The pair $(\Omega, \alpha)$ is
called a conformally symplectic structure on $N$ and it reduces
to a symplectic structure $\Omega$ whenever $\alpha =0$. Since
$\Omega$ is non-degenerate this is the same as $E=0$
\cite{kir76},\cite{LM}.
\begin{theorem}                   \label{prop10}
Let $\theta_k$ be a relative invariant so that $\varphi_k \neq constant$.
Then, for each $l$ the pair
\begin{equation}
   \Omega_{kl} \equiv \varphi_k \Omega_l \; ,\;\;\;
          \alpha_k \equiv d log \varphi_k              \label{okl}
\end{equation}
defines a conformally symplectic structure on $I \times M$
and an isomorphisms from $C^{\infty}(I \times M)$ into
vector fields on $I \times M$.
For a function $f$, the vector field $V_{klf}$ assigned by
(\ref{okl}) corresponds to the Hamiltonian vector field for the function
$f/ \varphi_k$ of the symplectic structure $\Omega_l$.
\end{theorem}
{\bf Proof:}
Since $\Omega_k$'s are closed $\Omega_{kl}$'s satisfy the conditions
(\ref{consym}) with $\alpha_k$ given as in (\ref{okl}).
To obtain the algebraic consequences we shall work with
contravariant objects.
The bi-vector dual to $\Omega_k$ can be computed to be
\begin{equation}
      P_k= - (  {\bf W}_k \cdot {\bf n}_k)^{-1}
       [ {\bf W}_k \cdot \nabla  \wedge  \partial_t +
     +( {\bf W}_k \times {\bf v} - \rho_{\varphi}^{-1}{\bf n}_k)
     \cdot \nabla \wedge \nabla ]               \label{bik}
\end{equation}
and one finds that the contravariant version of (\ref{consym}) is the
Jacobi structure defined by the pair 
\begin{equation}
   P_{kl} \equiv {1 \over \varphi_k} P_l \;,\;\;\;\;
   E_{kl} \equiv - {1 \over \varphi^{2}_k} P_l(d \varphi_k)   \label{fjac}
\end{equation}
which is conformally equivalent to $P_l$. The pair (\ref{fjac})
satisfy Eqs.(\ref{jjac}) and hence the brackets
\begin{eqnarray}
   \{ f,g \}_{kl} &=& P_{kl}(df \wedge dg) +  E_{kl}(fdg-gdf) \label{jb} \\
   &=& {1 \over \varphi_k} \{ f,g \}_l -
      {f \over \varphi^2_k} \{ \varphi_k,g \}_l +
          {g \over \varphi^2_k}  \{ \varphi_k,g \}_l
\end{eqnarray}
where $\{ \;,\; \}_l$ is the Poisson bracket defined by the bi-vector
(\ref{bik}), fulfill the Jacobi identity (\ref{fgh}) for each pair $(k,l)$.

The Jacobi structure also provides an isomorphism
between the Lie algebra of vector fields on $I \times M$ and the algebra of
functions on $I \times M$ with the local bracket (\ref{jb}).
If we let $V_{klf}$ denote the vector field corresponding to the
function $f$ defined by the Jacobi structure (\ref{fjac})
indexed by $(k,l)$, then we find
\begin{equation}
   V_{klf} = P_{kl}(df)+fE_{kl}
   = {1 \over \varphi_k} P_l(df)-{f \over \varphi^2_k} P_l(d \varphi_k)
   = P_l(d(f/ \varphi_k)) = V_{l(f/ \varphi_k)}
 \label{jacvec}       \end{equation}
which is the Hamiltonian vector field for the function $f/ \varphi_k$
defined by the Poisson bi-vector $P_l$.
Note that the vector field $E_{kl}$ is the Hamiltonian vector field
for the function $f=1$.
$\bullet$

Thus, in addition to the symplectic structures $\Omega_k$, we have
the families $\Omega_{kl}$ of conformally symplectic
structures on $I \times M$ that coincide with $\Omega_l$
on the level surfaces $\varphi_k =constant$ of Hamiltonian function
on which Eulerian conserved densities become Lagrangian invariants.
We therefore conclude
that the absolute invariance of the canonical one-form,
the degeneration of Eulerian conservation laws into Lagrangian invariants
and, the conformal equivalence of the local structures (\ref{okl})
to the symplectic structures $\Omega_l$ as well as of their
contravariant versions in Eqs.(\ref{fjac}) and (\ref{bik}) are
all the same. This verifies the statement of proposition (\ref{ppp}).

\section{Summary, discussions and conclusions}

We shall summarize the main constructions of the last two sections,
discuss the results and compare them with other works on helicity
conservations.
We shall indicate some generalizations as well as prospectives for
a dynamical interpretation of helicity invariants.

\subsection{Summary}

Given the frozen in fields $B$ and $\varphi$, the Eulerian
dynamical equations imply that $\partial_t+v$
is Hamiltonian with
$$\Omega = \omega + (i(v)(\omega) -d_M \varphi) \wedge dt \;.$$
The left invariant generators $W_k$ of particle
relabelling symmetries can be obtained from infinitesimal
Hamiltonian automorphisms of $\Omega$. The Hamiltonian structure
on $M$ of $W_k$'s is defined by the Nambu-Poisson bracket
$$\{ f,g \}_{\varphi}= (\rho_{\varphi})^{-1} \nabla \varphi \cdot
\nabla f \times \nabla g \; .$$
For each $k$, $i(W_k)(\mu_M)=-\omega_k$
is a closed two-form on $M$. These can be extended to exact,
degenerate two-forms
\begin{equation}
  \Theta_k=\omega_k+i(v)(\omega_k) \wedge dt=d \theta_k \;,\;\;
-\theta_k = \psi_k dt+ A_k   \label{tet}  \end{equation}
on $I \times M$ which are absolutely invariant under the flow
of $v$. The potential one-forms $\theta_k$ can be classified in
accordance with their orientation and invariance with respect
to the flow of $v$.
Those which are absolutely invariant give the Lagrangian
\begin{equation}
  {\partial {\cal H}_{kl} \over \partial t} +
    {\bf v} \cdot \nabla {\cal H}_{kl} =0
\label{ll}  \end{equation}
while the relatively invariant ones imply the Eulerian
\begin{equation}
  {\partial {\cal H}_{kl} \over \partial t}  + \nabla \cdot
  [ {\cal H}_{kl} {\bf v} - \rho_{\varphi} \varphi_k {\bf W}_l) ] =0 
\label{el}  \end{equation}
conservation laws for the generalized helicity densities
$$  {\cal H}_{kl} = \rho_{\varphi} {\bf a}_k \cdot {\bf W}_l
  = \rho_{\varphi} ( W_l (\lambda_k) - {\bf A}_k \cdot {\bf W}_l ) \;.$$
The densities satisfying Eqs.(\ref{ll}) and (\ref{el}) can be parametrized,
apart from the discrete parameters $k,l$, by functions in
$ker(\partial_t+v)$ and $C^{\infty}(I \times M)/ker(\partial_t+v)$,
respectively.
The invariant forms defining the conservation laws admit extentions
\begin{equation}
  \psi_k \mapsto \psi_k + \varphi_k  \; \Rightarrow \;
  \theta_k \mapsto \tilde{\theta}_k  \;,\;\;\;
  \Theta_k = d \theta_k \mapsto \Omega_k = - d \tilde{\theta}_k  \;.
\end{equation}
to symplectic forms on $I \times M$. The conservation laws are
expressed by the identity $d(\tilde{\theta}_k \wedge \Omega_l) +
\Omega_k \wedge \Omega_l \equiv 0$.
When $\tilde{\theta}_k $ is relatively invariant, this gives an
Eulerian conservation law and the pair
$\varphi_k \Omega_l \; ,\; \alpha_k \equiv d log \varphi_k $
defines a conformally symplectic structure.
On the hypersurfaces $\varphi_k =constant$ these degenerate into
a Lagrangian conservation law and the symplectic structure $\Omega_l$,
respectively.

\subsection{Discussions and prospectives}

We constructed helicity conservation laws from the invariant
differential forms associated with the particle relabelling symmetries.
This can be continued by introducing new families of invariant
forms. For example,
since $W_k$'s commute with $\partial_t+v$ we can construct 
invariant two-forms by taking Lie derivatives of $\Omega$ with
respect to $W_k$'s. This we can
compute using ${\cal L}_{U_k} (\Omega) =0$ which follows from
the fact that $U_k$'s are Hamiltonian.
Using the decomposition (\ref{uvw}) and the Hamilton's equations
for $\partial_t+v$ we get
\begin{equation}
  {\cal L}_{W_k} (\Omega) = d \varphi \wedge d \xi_k
                       \equiv \Sigma_k               \label{sigma}
\end{equation}
which means that $\Omega$ is a relative invariant for
$W_k, \; \forall k$. The degenerate two-forms $\Sigma_k$ are absolute
invariants for $\partial_t+v$ and are exact $ \Sigma_k =  d \sigma_k $.
The gauge group $C^{\infty}(I \times M)$
also enters into the definition of $\Sigma_k$'s
and one can proceed as above to construct helicity type
conservation laws expressed as closure of three-forms
$\sigma_k \wedge \Sigma_l, \theta_k \wedge \Sigma_l$ and
$\sigma_k \wedge \Theta_l$.

The helicity invariant which is first discovered in \cite{skr66}
have been studied in Refs.
\cite{mor61}-\cite{yah95},\cite{sal88},\cite{she92},\cite{pad98}
in the context of Noether theorems.
The ergodic and topological interpretations of helicity type
invariants for three-dimensional flows were introduced and
investigated in Refs.
\cite{mof69}-\cite{fre88},\cite{arn74},\cite{khc89}.
A relation between infinite families of (magnetic) helicity
invariants and magnetic surfaces has been remarked in Ref. \cite{mof92}.
The present construction inherits geometric objects for investigation
of coadjoint orbit invariants. The framework can also be exploited
for studying the interplay between these invariants each of which
has been discussed separately in various contexts.
For example, it offers $W_k$'s for linking numbers,
$\omega_k$'s for Hopf invariants, $\Theta_k$'s for Novikov type
invariants and the Godbillon-Vey type invariants arises
from the nilpotent generators of volume preserving diffeomorphisms.

We presented the kinematical aspects of helicity invariants
in the geometric language of Jacobi structures.
The results of proposition (\ref{prop10}) are suggestive for
further investigation, in the context of divergence-free vector fields,
of the relations between Hamiltonian vector fields isomorphic to a local
Lie algebra and Hamiltonian vector fields isomorphic to a Poisson
bracket algebra.
There is also a dynamical content of helicity invariants
which will be presented in a forthcoming article.
This is connected with Liouville structures
on $I \times M$. A Liouville structure can be defined by
a one-form together with an action of the multiplicative
group $R_*$ of non-zero real numbers \cite{LM}.
In the present context the one-form is precisely the
canonical one-form $\theta$ of the symplectic structure.
The action of $R_*$ is generated by the vector field which
is dual to the three-form $\theta \wedge \Omega$ with
respect to the symplectic volume. The dynamical properties
of the fluid, such as viscosity, are implicit in this
generator. Its divergence gives the evolution equation
for the helicity density. This reduces to a conservation
law for inviscid flows. Its action by Lie derivative
corresponds to scaling transformations.

\subsection{Conclusions}

The symplectic structure on $I \times M$ provided us not only
the way to construct infinitely many helicity type conservation
laws associated with the Lie algebra of divergence free vector fields
but also a kinematical interpretation of them with the Lie
algebraic structures on function spaces over finite dimensional
manifold $I \times M$.
In Ref. \cite{kup87a} a similar interpretation with the local Lie
algebras of Hamiltonian systems of hydrodynamic type, as introduced in
Refs. \cite{kum77} and \cite{kum77a}, was described in the general
framework of infinite dimensional Lie algebras.

The type of conservation laws associated with the particle
relabelling symmetries and, in particular, the construction
of infinitely many helicity type invariants for three dimensional
flows seem to be much related to and rely on the conformal
properties of the space of trajectories. The present framework
incorporates the conformal transformations, which are not contained
in $Diff_{vol}(M)$, into the study of kinematical invariants
in connection with the algebraic structures on function spaces.
We showed that Jacobi structures being associated with
local Lie algebras provides a framework for investigation of properties
which can not be obtained from the geometry of $Diff_{vol}(M)$.


\begin{thebibliography}{99}
\bibitem{arn66}
       V. I. Arnold,
       Sur la g\'eom\'etrie diff\'erentielle des groupes de Lie
       dimension infinite et ses applications \`a l'hydrodynamique
       des fluides parfaits,
       Ann. Inst. Fourier 16 (1966) 319-361.
\bibitem{mwe74}
       J. E. Marsden and A. Weinstein,
        Reduction of symplectic manifolds with symmetry,
             Rep. Math. Phys. 5 (1974) 121-130.
\bibitem{mwe83}
          J. E. Marsden and A. Weinstein,
          Coadjoint orbits, vortices, and Clebsch variables
          for incompressible fluids,
          Physica D 7 (1983) 305-323.
\bibitem{ARN} 
      V. I. Arnold, Mathematical Methods of Classical
 Mechanics, Graduate Text in Mathematics, Vol: 60, Second Edition,
   (Springer, Berlin, 1989)
\bibitem{ebm70} 
           D. Ebin and J. E. Marsden,
           Groups of diffeomorphisms and the motion of an incompressible
           fluid,
             Ann. of Math. 92 (1970) 102-163.
\bibitem{djr86} 
      D. D. Holm, J. E. Marsden and T. S. Ratiu,
      The Hamiltonian structure of continuum mechanics in material,
      spatial and convective representations,
    S\'eminaire de Math\'ematiques Sup\'erieurs,
    Les Presses de L'Universit\'e de Montr\`eal, 100 (1986) 11-122.
\bibitem{MR} 
        J. E. Marsden and T. Ratiu, Introduction to Symmetry
                and Mechanics, Texts in Applied Mathematics, Vol.17,
                  (Springer, Berlin, 1994).
\bibitem{khc89}
          B. A. Khesin and Yu. V. Chekanov,
          Invariants of the Euler equations for ideal or barotropic
          hydrodynamics and superconductivity in D dimensions,
          Physica D 40 (1989) 119-131.
\bibitem{ark92}
          V. I. Arnold and B. A. Khesin,
          Topological methods in hydrodynamics,
                Ann. Rev. Fluid Mech. 24 (1992) 145-166.
\bibitem{arn74}
         V. I. Arnold,
         The asymptotic Hopf invariant and its
    applications, in: Proc. Conf. on Diff. Eqs. 1973, Yerevan,
    English translation: Sel. Math. Sov. 5(4) (1986) 327-345.
\bibitem{dez83}
        A. A. Dezin,
        Invariant forms and some structure properties of the
 Euler equations of hydrodynamics, Z. Anal. Anwend 2 (1983) 401. (in Russian)
\bibitem{hok83}
         D. D. Holm and B. A. Kupershmidt,
        Poisson structures of superconductors,
                  Phys. Lett. A 93 (1983) 177-181.
\bibitem{gik92}
         V. L. Ginzburg and B. A. Khesin, Topology of steady fluid flows,
       in: Topological Aspects of the Dynamics of Fluids and Plasmas,
                  NATO-ASI Series E, vol. 218, eds. H. K. Moffatt,
                  G. M. Zaslavsky, P. Comte and M. Tabor,
                  (Kluwer, Dordrecht, 1992).
\bibitem{ser84}
        D. Serre,
        Invariants et d\'eg\'en\'erescence symplectique de l'\'equation
        d'Euler des fluides parfaits incompressibles,
         C. R. Acad. Sci. Paris Ser. A 298 (1984) 349-352.
\bibitem{FEYN}
         R. P. Feynman, Statistical Mechanics,
             (Benjamin, New York, 1972)
\bibitem{hmr83} 
         D. D. Holm, J. E. Marsden, T. Ratiu and A. Weinstein,
         Nonlinear stability conditions and a priori estimates
         for barotropic hydrodynamics,
                  Phys. Lett. A 98 (1983) 15-21.
\bibitem{vid78}
      S. V. Vishik and F. V. Dolzanskii,
      Analogs of the Euler-Lagrange equations and magnetohydrodynamics
      equations connected with Lie groups,
      Sov. Math. Dokl. 19 (1978) 149-153.
\bibitem{hok83a}
         D. D. Holm and B. A. Kupershmidt,
        Poisson brackets and Clebsch representations for
        magnetohydrodynamics, multifluid plasmas, and elasticity,
                  Physica D 6 (1983) 347-363.
\bibitem{mor98}
            P. J. Morrison,
            Hamiltonian description of ideal fluid,
            Rev. Mod. Phys. 70 (1998) 467-521.
\bibitem{vcs90} 
         G. K. Vallis, G. F. Carnevale and T. G. Shepherd,
      A natural method for finding stable states of Hamiltonian systems, 
  in: Topological Fluid Dynamics, Proc. IUTAM Symposium, Cambridge, 1989,
                           eds. H. K. Moffatt and A. Tsinober
                          (Cambridge University Press, Cambridge, 1990).
\bibitem{hg97} 
          H. G\"umral,
          Lagrangian description, symplectic structure and
          invariants of 3D fluid flow,
          Phys. Lett. A232 (1997) 417-424.
\bibitem{hg98} 
          H. G\"umral,
           Kinematical symmetries of 3D incompressible fluids,
          Physica D (1998)
\bibitem{pad98} 
          N. S. Padhye,
          Topics in Lagrangian and Hamiltonian fluid mechanics:
          Relabelling symmetry and ion-acoustic wave stability,
          Dissertation, Institute for Fusion Studies,
          University of Texas, Austin, 1998.
\bibitem{swh68}
        R. L. Seliger and G. B. Whitham,
         Variational principles in continuum mechanics,
        Proc. Roy. Soc. A 305 (1968) 1-25.
\bibitem{sta90}
            J. T. Stuart and M. Tabor,
            The Lagrangian picture of fluid motion,
            Phil. Trans. R. Soc. Lond. A 333 (1990) 263.
\bibitem{sal88}
      R. Salmon,
      Hamiltonian fluid mechanics,
      Ann. Rev. Fluid Mech. 20 (1988) 225-256.
\bibitem{frv90}
          S. Friedlander and M. M. Vishik,
                 Lax pair formulation for the Euler equation,
                        Phys. Lett. A 148 (1990) 313-319.
\bibitem{CM} 
       P. Chernoff and J. E. Marsden,
       Properties of infinite
               Dimensional Hamiltonian Systems,
       Lecture Notes in Mathematics Vol. 425 (Springer, Berlin, 1974)
\bibitem{OLV} 
         P. J. Olver,
         {\it Applications of Lie Groups to Differential Equations},
         Springer, Berlin, 1986.
\bibitem{SAUN}
          D. J. Saunders, Geometry of Jet Bundles, London Mathematical
   Society Lecture Note Series 142 (Cambridge University Press, 1989). 
\bibitem{VAR}
      V. S. Varadarajan, {\it Lie Groups, Lie Algebras and
Their Representations}, Graduate Texts in Mathematics Vol. 102
                          (Springer, NewYork, 1984) page:411.
\bibitem{POST}
      M. M. Postnikov, {\it Lectures in Geometry, semester V,
Lie Groups and Lie Algebras} (Mir, Moscow, 1986) Lecture 17.
\bibitem{tab90}
     S. L. Tabashnikov,
     Two remarks on asymptotic Hopf invariants,
     Func. Anal. Appl. 24(1) (1990) 74-75.
\bibitem{gun93} 
         H. G\"{u}mral and Y. Nutku,
          Poisson structures of dynamical
                     systems with three degrees of freedom, 
             J. Math. Phys. {\bf 34} (1993) 5691-5723
\bibitem{hag98} 
          H. G\"umral,
           A time-extended Hamiltonian formalism,
          Phys. Lett. A (1998)
\bibitem{she92}
       T. G. Shepherd,
       Extremal properties and the Hamiltonian
        structure of the Euler equations,
       in: Topological Aspects of the Dynamics of Fluids and Plasmas,
                  NATO-ASI Series E, vol. 218, eds. H. K. Moffatt,
                  G. M. Zaslavsky, P. Comte and M. Tabor,
                  (Kluwer, Dordrecht, 1992).
\bibitem{AMR}
       Abraham, R., Marsden, J.E., and Ratiu, T., {\it  Manifolds, tensor
analysis, and applications}, Addison Wesley, Reading, 1983.
\bibitem{LM} 
        P. Libermann and C.-M. Marle,
         {\it Symplectic Geometry and Analytical Mechanics},
         D. Reidel Publishing Company, Dordrecht, 1987.
\bibitem{pam96} 
          N. S. Padhye and P. J. Morrison,
          Relabelling symmetries in hydrodynamics and
          magnetohydrodynamics,
          Plasma Physics Reports, 22 (1996) 869-877.
\bibitem{pam96a}
          N. S. Padhye and P. J. Morrison,
         Fluid element relabelling symmetry,
         Phys. Lett. A 219 (1996) 287-292.
\bibitem{ahm86}
        H. D. I. Abarbanel, D. D. Holm, J. E. Marsden and T. S. Ratiu,
        Nonlinear stability analysis of stratified fluid equilibria,
            Phil. Trans. R. Soc. Lond. A 318 (1986) 349-409.
\bibitem{ah87} 
       H. D. I. Abarbanel and D. D. Holm,
   Nonlinear stability analysis of inviscid flows in three dimensions:
   incompressible fluids and barotropic fluids,
                Phys. Fluids 30 (1987) 3369-3382.
\bibitem{kir76}
          Kirillov, A. A.,
          Local Lie algebras,
          Russian Math. Surveys 31(4) (1976) 55-75.
\bibitem{lic78}
       A. Lichnerowicz, 
      Les vari\'et\'es de Jacobi et leurs alg\`ebres de Lie associ\'ees,
       J. Maths. Pures. Appl. 57 (1978), 453-488.
\bibitem{skr66}
       M. Steenbeck, F. Krause and K. H. R\"adler,
       A calculation of mean electromotive force in an electrically
       conducting fluid in turbulent motion under the influence
       of coriolis forces,
       Z. Naturforsch 21a (1966) 369.
\bibitem{mor61}
       J. J. Moreau,
        Constantes d'un ilet tourbillonnaire en fluid parpait
        barotrope, C. R. Acad. Sci. Paris 252 (1961) 2810.
\bibitem{mor77}
       J. J. Moreau,
       Sur les int\'egrales premi\`eres de la dynamique d'un
       fluid parpait barotrope et le th\'eor\`eme de
       Helmholtz-Kelvin, S\'eminaire d'Analyse Convex,
       Montpellier, Expos\'e no.7 1977.
\bibitem{cal63}
       M. G. Calkin,
        An action principle for MHD,
        Can. J. Phys. 41 (1963) 2241-2251.
\bibitem{yah95}
         A. Yahalom,
         Helicity conservation via the Noether theorem,
                               J. Math. Phys. {\bf 36} (1995) 1324-1327
\bibitem{mof69}
        H. K. Moffatt,
        The degree of knottedness of tangled vortex lines,
        J. Fluid Mech. 35 (1969) 117.
\bibitem{kum80}
         E. A. Kuznetsov and A. V. Mikhailov,
         On topological meaning of canonical Clebsch variables,
                 Phys. Lett. A 77 (1980) 37-38.
\bibitem{fre88}
         M. H. Freedman,
         A note on topology and magnetic energy in incompressible
         perpectly conducting fluids,
         J. Fluid Mech. 194 (1988) 549-551.
\bibitem{mof92}
        H. K. Moffatt, Topological (as opposed to the analytical)
                    approach to fluid and plasma flow problems, 
       in: Topological Aspects of the Dynamics of Fluids and Plasmas,
                  NATO-ASI Series E, vol. 218, eds. H. K. Moffatt,
                  G. M. Zaslavsky, P. Comte and M. Tabor,
                  (Kluwer, Dordrecht, 1992).
\bibitem{kup87a}
           B. A. Kupershmidt,
           Hydrodynamical Poisson brackets and local Lie algebras,
              Phys. Lett. A {\bf 121} (1987) 167-174.                          
\bibitem{kum77} 
           B. A. Kupershmidt and Yu. I. Manin,
           Long-wave equations with free boundaries. I. conservation laws,
           Func. Anal. Appl. 11 (1977) 188-197.
\bibitem{kum77a} 
           B. A. Kupershmidt and Yu. I. Manin,
           Equations of long waves with a free surface.
           II. Hamiltonian structure and higher equations,
           Func. Anal. Appl. 12 (1978) 20-29.
\end{thebibliography}
\end{document}